\documentclass[conference]{IEEEtran}
\ifCLASSINFOpdf
\else
\fi
\usepackage{graphics} % for pdf, bitmapped graphics files
\usepackage{amsmath} % assumes amsmath package installed
\usepackage{lipsum}% http://ctan.org/pkg/lipsum
\usepackage{graphicx}
\usepackage{array}

\begin{document}
\title{Microgrid Control Using Remote Controller Hardware-in-the-Loop Over the Internet}

\author{\IEEEauthorblockN{Amir Valibeygi\\ and Raymond A de Callafon}
\IEEEauthorblockA{Mechanical and Aerospace Engineering\\
University of California San Diego\\
La Jolla, California\\
avalibey@eng.ucsd.edu}
\and
\IEEEauthorblockN{Mark Stanovich, Michael Sloderbeck \\ Karl Schoder, James Langston\\ Isaac Leonard, Sourindu Chatterjee}
\IEEEauthorblockA{Center for Advanced Power Systems (CAPS)\\Florida State University\\Tallahassee, Florida}
\and
\IEEEauthorblockN{Rick Meeker}
\IEEEauthorblockA{Nhu Energy Inc.\\ Tallahassee, Florida}
}


%Telephone: (800) 555--1212, Fax: (888) 555--1212}


\maketitle

\begin{abstract}
A centralized microgrid power management and control system is developed and tested with a Hardware-In-the-Loop (HIL) Real-Time Digital Simulator (RTDS) model of an existing microgrid that communicates in real-time with the controller over the Internet. The controller leverages Phasor Measurement Units (PMUs) for measuring power flow and adjusting inverter power references in real-time. The control objectives are power and State-of-Charge (SoC) control, subject to inverter power amplitude and rate limits and communication constraints. The controller incorporates units of power control, SoC control, and adaptive reference scheduling to achieve seamless microgrid operation. Real-time, over the internet, hardware-in-the-loop tests between the controller and the simulator are realized and indicate stability and performance of the microgrid control system.\end{abstract}


% \fi
\IEEEpeerreviewmaketitle

\section{Introduction}
Distributed Generation (DG) is reaching unprecedented levels of penetration in the global power industry. This has been realized as more economic and efficient Distributed Energy Resources (DER) have become available \cite{2016report}. A microgrid is a group of local DERs, storage units, and loads that are able to operate in both connected and isolated modes from the main electricity grid \cite{2014trends}. In the connected mode, microgrids are connected to the main grid at the Point of Common Coupling (PCC) and may provide part, or all of the power demand of their local facility; hence reducing power required from the main grid. Further, excess power generation may be stored or sold to the main grid. 
% Microgrids can switch to the islanded mode to survive a period of power outage and independently serve the local demand. Otherwise, microgrids may be permanently operating in the islanded mode because they are located remotely from the main electricity grid.

In grid-connected mode of operation, the microgrid can regulate the amount of active and reactive power exchange with the main grid. Numerous planning, energy management, and control strategies have been devised to manage the flow of power between microgrid and the main grid, in order to maximize economic efficiency and maintain reliability and availability \cite{callafon2016scheduling,ahn2013power,zhang2013robust,shahidehpour2013cutting}.

Microgrid operations and controls can be significantly improved by utilizing Phasor Measurement Units (PMUs). PMUs provide synchronized real-time measurements (synchrophasors) of various quantities at multiple points across the microgrid \cite{de2010synchronized}, \cite{konakalla2016optimal}. 
%With the increased availability of PMUs for use in power technology applications, high quality automation, state estimation, and event detection can be introduced to power systems  
%Although PMUs provide an unprecedented level of fidelity in electric power system measurements for use in control, they also pose communication and computational challenges due to the massive amount of data they generate. PMUs can be utilized at certain points within the microgrid to provide exact and frequent measurements of voltage and current phasors and therefore compute power and frequency. 
Control strategies can be implemented by feedback loops that use such PMU measurements. Furthermore, any undesired behavior in the microgrid including faults, power fluctuations at the PCC due to the varying demand, etc., could be sensed and sent to a central controller that takes appropriate control actions accordingly to ensure quality, reliability, and economic operation \cite{uhlen2012wide}. For large electricity consumers in market, total electricity charge is calculated based on both the consumed energy and peak power demand during a period. Therefore, it is imperative to control large power spikes due to load switchings in such big facilities. This can be realized by means of a feedback control system using PMU data.

%\smallskip
%Various techniques have been adopted for microgrid energy management and control in the grid-connected mode \cite{2014trends}. Most techniques utilize droop control schemes to control power in a decentralized manner \cite{gao2008control,de2007voltage}. Such controllers are completely local and work well as long as the balance between generation and consumption is maintained. However, a centralized control layer can be added on top of droop controllers to control power flow at the PCC or implement other scheduling scenarios.

In this work, power management and control is implemented to manage power flow at the PCC of a grid-connected microgrid. In the first stage of this work, a real-time simulation of a microgrid equipped with inverter-interfaced PV generation and battery storage was developed for developing a microgrid power management and control algorithm. A Controller Hardware-In-the-Loop (CHIL) setup is established wherein a RTDS-based microgrid simulator (CAPS, FSU) and controller (SyGMA Lab, UCSD) communicate over the Internet. The controller performs computations in real-time and sends control commands to the inverter that is part of the microgrid model. The long-distance communication testbed has enabled two research entities with different expertise and resources, separated by over 2000 miles, to succesfully collaborate throughout the development and testing process. This real-time control platform over the Internet has extensive potential applications for various experiments and validation tests in the power industry \cite{CAPS2016}. 

%The inverter has inherent ramp rate limitation as well as maximum power saturation limit. The microgrid comprises several loads with different degrees of importance. 
The objective of the control system is to track the power reference for PCC and simultaneously maintain the State-of-Charge (SoC) of the battery within an acceptable range while conforming to system (e.g., inverter and communications delay) limitations. This two-fold objective is achieved by utilizing a centralized cascaded control system. The controller incorporates a fast inner loop that aims at power control and a slower event-triggered outer loop for SoC control. A reference calculation module is implemented in the controller that estimates the demand profile (disturbance) and sets the power reference based on the estimation. 

\begin{figure}[thpb]
    \centering
    \includegraphics[trim=.8in 0 .5in 0,clip,width=8.5cm]{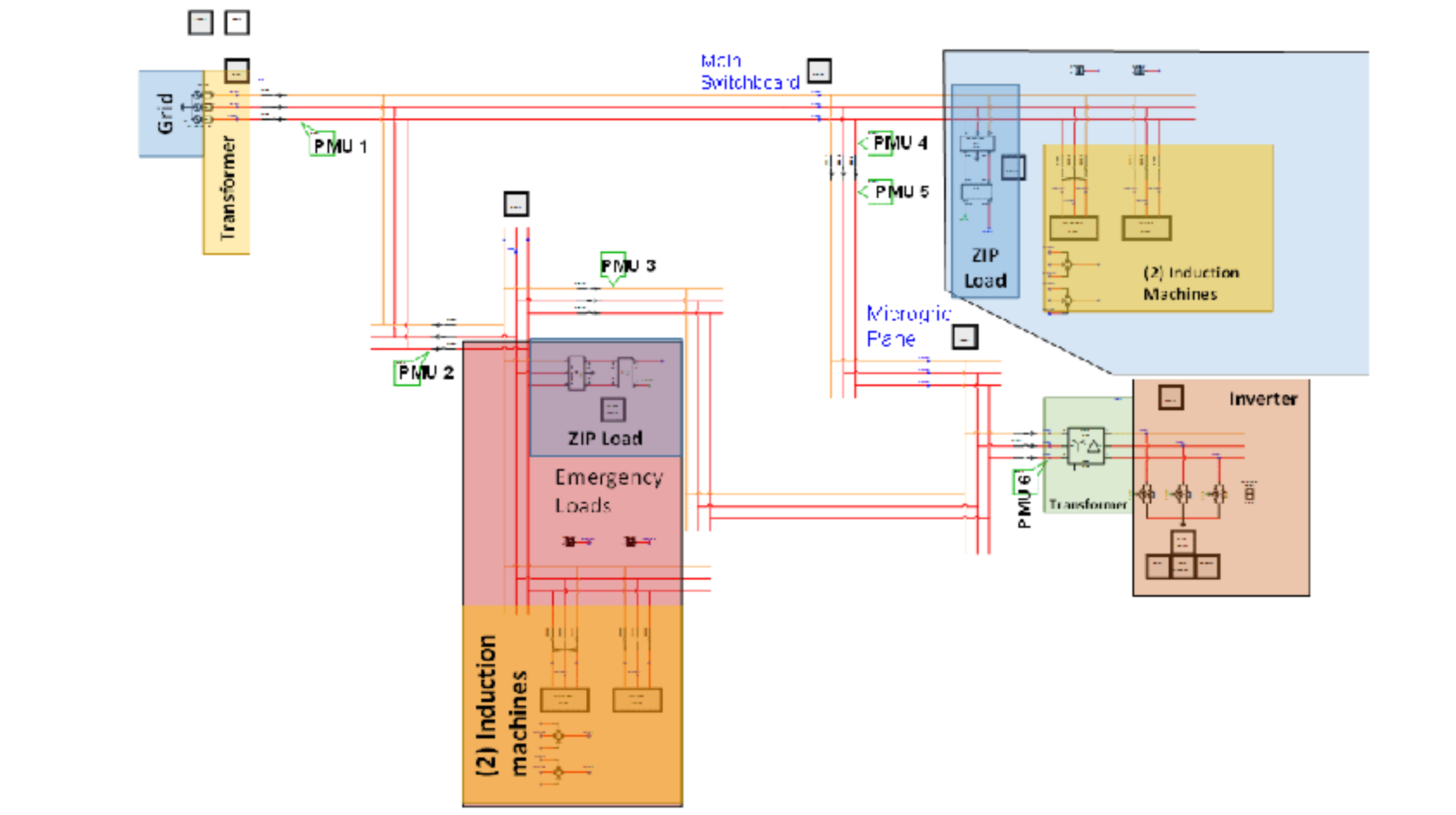}
    \caption{Overview of RTDS microgrid model, emergency and non-emergency loads, PMUs, and transformers}
    \label{MicrogridModelIllustration}
\end{figure}
%In the following, Section II describes the microgrid model and real-time simulation, as well as the control algorithm developed for the microgird. Section III discusses the CHIL environment, controller, and long-distance communication. Section IV presents microgrid control tests and their results that capture multiple scenarios and control modes, and Section V provides concluding remarks.
%%%%%%%%%%%%%%%%%%%%%%%%%%%%%%%%%%%%%%%%%%%%%%%%%%%%%%%%%%%%%%%%%%%%%%%%%%%%%%%%

\section{Verbose System Model and Control Strategy}

%In this section, developments for this work are described including: the microgrid model, its implementation for real-time simulation, microgrid system identification, and the controller.

\subsection{Microgrid Model and Simulation}

A microgrid model was developed in this project to capture salient characteristics of an existing hospital electrical system and perform real-time simulation with RTDS Technologies hardware and software \cite{RtdsTechnologies}. This model and simulation is primarily used for de-risking and development of controls for planned hardware additions to the hospital electrical system including PV and batteries. A high-level illustration of microgrid model in the RTDS design environment, along with annotations, is shown in Figure~\ref{MicrogridModelIllustration}. The microgrid model has loads which can be categorized as non-emergency and emergency. The emergency loads draw much less power than the non-emergency loads and can therefore be powered solely by the planned hardware installation. The emergency and non-emergency loads each consist of a constant impedance-current-power (ZIP) load and two induction machines. The grid interconnection is modeled using an infinite source and transformer equivalent impedance. The modeled planned additions to the microgrid include 6 PMUs, a PV array, an inverter, and a battery. The inverter and battery storage are rated at 250~kW/250~kVar and 250~kW/1~MWh, respectively.

TCP/IP Modbus and C37.118 data communication is implemented in the real-time simulation. 
The model includes 6 PMUs that send C37.118 messages providing measurements throughout the microgrid. The simulated inverter provides a Modbus TCP/IP interface, which is the communication channel for controlling real and reactive power and information including battery SoC and PV power generation.

\subsection{Microgrid Dynamic Model Estimation}

The dynamics of the central microgrid controller is developed based on only an approximate model of the detailed microgrid dynamics captured in the RTDS model. The approximate model is found through system identification tests, which were performed to model the relationship between the power flows at the inverter and PCC. To achieve this, step tests are conducted with the inputs being active and reactive power at the inverter and the outputs active and reactive power flow measured at the PCC by PMU 1. A two-input-two-output discrete-time model $\hat{G}$ between the inverter and the PCC is identified through a step-based realization algorithm \cite{callafon2013step}.
\begin{align}
&\hat{G}(q)=\begin{bmatrix}
\hat{G}_{11}(q)&\hat{G}_{12}(q)\\\hat{G}_{21}(q)&\hat{G}_{22}(q)
\end{bmatrix}
\label{eq1}
\end{align}
The identified model is used in the next section to design controllers and to estimate microgrid's power demand.

\subsection{Control Strategy and Implementation}

 \begin{figure}[thpb]
      \centering
     \includegraphics[width=8cm]{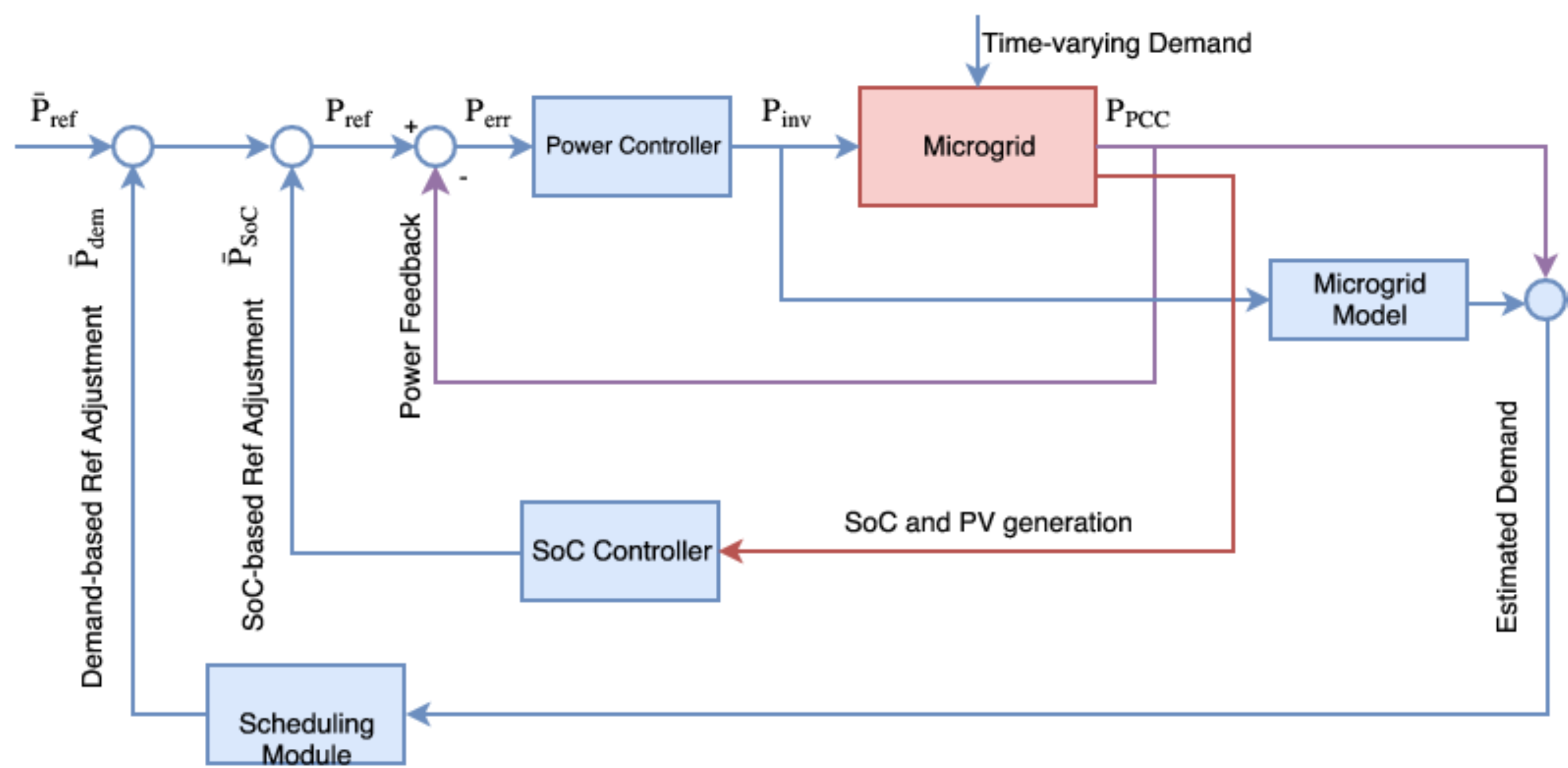}
      \caption{Cascaded controller block diagram consisting fast power control loop, slow SoC control loop, and reference adjustment module}
      \label{ControllerBlockDiagram}
   \end{figure}

The microgrid supervisory control system is a central controller that gathers PMU data as well as battery and PV generation data as input data. The collected information is processed and appropriate control commands (inverter active and reactive power reference) are computed and sent back as demand signals to the inverter via the Internet.

The goal of the proposed control system is to track a reference for active and reactive power at PCC, as long as SoC of the battery is within its acceptable range. The reference could be set either adaptively by computing a rate-limited tracking of demand trend estimate or set manually by an operator. Due to the limited inverter power, the adaptive reference allows the implementation of power peak shaving at different demand levels. On the other hand, the controller should also be able to follow a user-defined reference when requested allowing the suppression of short time, small magnitude demand variations. In the case that SoC drifts outside its acceptable range, the outer loop is triggered to refine the reference and recover the SoC until it reaches an acceptable level. The controller comprises three main units of power control, SoC control, and power scheduling which are described in more details below.
\section{Control Algorithms}

\subsection{Power Control}
Controlling power flow at the PCC is a primary objective of the microgrid central control system. Power control loop is the core control unit in the control system and is also the fastest loop. The input of this control unit is the inverter power error (adjusted by SoC control loop and reference calculation unit).
\begin{align}\nonumber
&P_{err}=P_{ref}-P_{PCC}\\
&P_{ref}=\bar{P}_{ref}+\bar{P}_{dem}+\bar{P}_{SoC}
\end{align}
where $\bar{P}_{dem}$ and $\bar{P}_{SoC}$ are refinements due to demand following and SoC control respectively. The controller incorporates proportional, integral, and derivative control actions. 
\begin{align}\nonumber
P_{inv}(k)=&K_P(P_{err}(k)-P_{PB})\\&+K_Ix_I(k)+K_D(x_D(k)-P_{DB})
\label{eq3}
\end{align}
where $x_I$ and $x_D$ are integrator and derivative states in the controller and $P_{PB}$ and $P_{DB}$ are bias terms to avoid proportional and derivative jumps and ensure bump-less transfer. The integrator state in the controller is updated as
\begin{align}\nonumber
x_{I}(k)=T_sP_{err}(k)+x_{I}(k-1)
\end{align}
and the output of the controller $P_{inv}(k)$ is the inverter's power references and is amplitude- and rate-limited to ensure it conforms with the inverter's limitations. As a result of this output limitation, integrator windup effect may occur in the integrator state. To avoid this, the computed control command $P_{inv}(k)$ is compared with amplitude and rate limits of the inverter and if the limits are violated, maximum possible power is commanded instead of the computed value and using \eqref{eq3}, the integrator state is updated by the following correction. 
\begin{align}\nonumber
x_I(k)=&[P_{inv,max}(k)-K_P(P_{err}(k)-P_{PB})\\&-K_D(x_D(k)-P_{DB})]/K_I
\end{align} 

\subsection{SoC Control}
While power control is the main objective of the control system, this task should be accomplished with the consideration of SoC level of the battery. Battery SoC should ideally be kept close to a nominal value; however, slow dynamics of SoC variation as well as the communication constraints motivate us to take an event-triggered approach to the SoC control loop. Therefore, this loop will remain inactive in the vicinity of the nominal SoC value (dead zone) where full control authority is dedicated to power control loop. If SoC drifts too far from its reference value, SoC compensation takes action to recover SoC and adjusting the power reference until it is brought back inside the dead zone. This reference correction would be helpful especially when average PV generation drops at night time or during cloudy days.% 
\begin{gather*}\bar{P}_{SoC}=
\begin{cases}
f(SoC,P_{PV})& \text{for } 20\le SoC < 30 \ \\
0 & \text{for } 30\le SoC\le 80\  \\
 f(SoC,P_{PV})& \text{for } 80 < SoC\le 90\    \end{cases} \end{gather*}%
This control loop involves proportional as well as integral control actions similar to \eqref{eq3}. Due to slow dynamics of the loop, integration windup limits are set to avoid large integrator accumulation and achieve steady-state error correction.

\begin{figure*}[thpb]
    \centering
    \includegraphics[width=14.5cm]{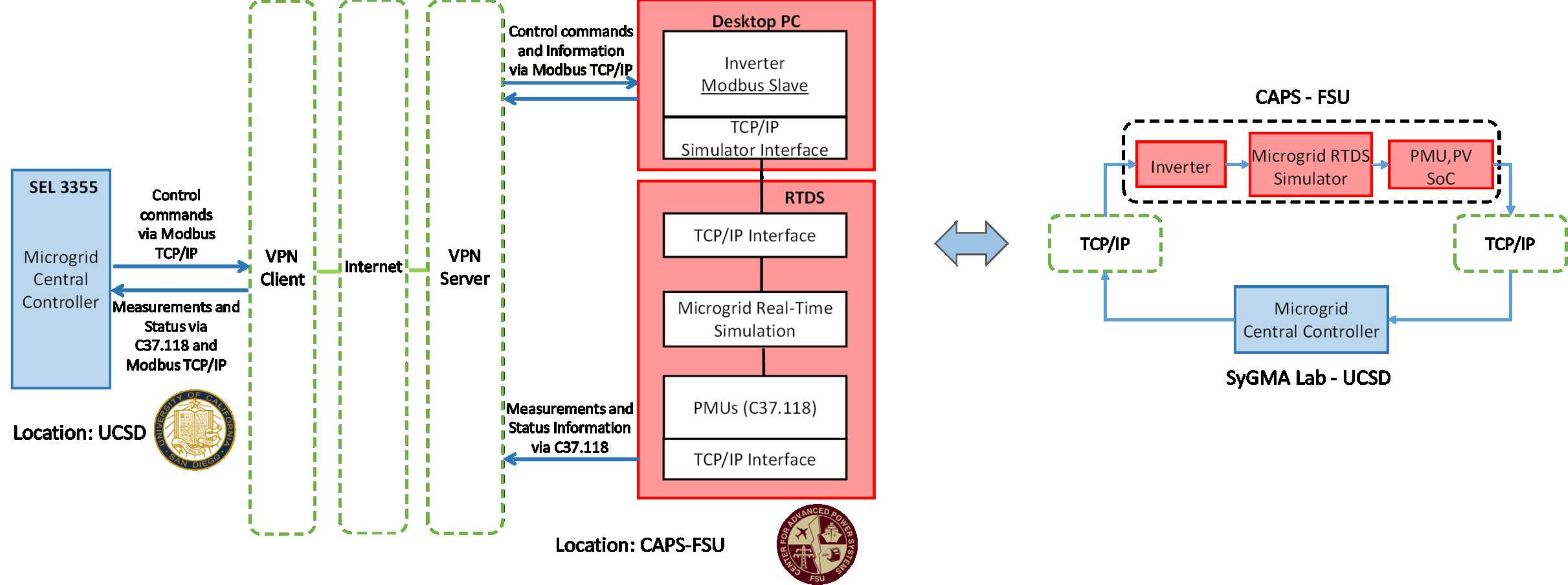}
    \caption{CHIL Setup with microgrid controller at UC San Diego and microgrid simulator at FSU communicating in real time over the Internet}
    \label{ChilSetup}
\end{figure*}

\subsection{Demand Estimation and Power Scheduling}
As depicted in fig. \ref{ControllerBlockDiagram}, this unit is in charge of providing power reference value based on either manual power reference or adaptive demand following and smoothening requirement depending on the operator's decision. The first strategy fits facilities with more frequent short-term load variations and constant long-term average while the latter suits those with low frequency power variations and variable average demand. In the latter case, in order to adapt the controller to the time-varying average demand, the demand is first estimated and then a rate-limited filtering of it is used as the adaptive power reference.  An estimation of instantaneous load demand at the PCC is computed by adding measured power at the PCC and inverter power filtered by the system model. 
\begin{align}\nonumber
\hat{P}_{dem}(k)= & P_{PCC}(k)+\\ & \hat{G}_{11}(q) P_{inv}(k-1)+\hat{G}_{12}(q) Q_{inv}(k-1)
\end{align} 
where $\hat{P}_{dem}(k)$ is the estimated microgrid demand, $P_{PCC}(k)$ is the measured power flow by the PMU at PCC, $\hat{G}_{11}(q)$ and $\hat{G}_{12}(q)$ are the identified system models from \eqref{eq1}, and $P_{inv}$ and $Q_{inv}$ are the inverter's control inputs. Next, a rate-limited version of this estimation is used as the adaptive power reference. If the rate limit parameter is $R$, we define the rate $r$ for the demand signal $\hat{P}_{dem}$ as
\begin{align}
r(k)=\frac{\hat{P}_{dem}(k)-\bar{P}_{dem}(k-1)}{T_s}
\end{align}
and the output rate limitation is performed as
\begin{gather*}\bar{P}_{dem}(k)=
\begin{cases}
T_sR+\bar{P}_{dem}(k-1)& \text{for } r>R \ \\
\hat{P}_{dem}(k) & \text{for } -R \le r \le R\  \\
 -T_sR+\bar{P}_{dem}(k-1)& \text{for } r < -R\    \end{cases} \end{gather*}
This enables the microgrid to automatically change its reference and achieve power control at different load levels. The output of this unit will then be adjusted by the output of the SoC control unit and later compared with the output of the system. Similar expression holds for reactive power at the PCC.

Due to the electric microgrid network between the generation units and the PCC, some active-reactive power coupling exists between power injected by the DG (inverter) and the measurement at the PCC. Therefore, in order to design independent control loops for controlling active and reactive power at the PCC, power decoupling is performed in the controller. 
%Besides these main units, additional measures are taken in the development of the controller to ensure smooth and safe transition of all electrical variables and to meet the plant’s inherent limitations. These measures include:

%\begin{itemize}
%\item Update rate limitation on inverter's generated power
%\item Limitation on communication frequency between the main controller and the plant
%\item SoC boundary dead zone to avoid SoC chattering
%\end{itemize}

\section{Implementation}
The real-time simulation and control facilities used for this project are located on opposite sides of the United States. The RTDS is located at the Center for Advanced Power Systems (CAPS), Florida State University. The digital controller is located at the Synchro-phasor Grid Monitoring and Automation (SyGMA) lab, San Diego Supercomputer Center, UC San Diego. A CHIL setup is created wherein the simulator and controller communicate over a virtual private network (VPN) as illustrated in Figure~\ref{ChilSetup}. Data from RTDS is communicated to the controller via TCP/IP at the rate of 10~Hz. The communicated data items are shown in Table~\ref{Tab:CommData}. PMU communication adheres to the IEEE C37.118 standard, which is the common IEEE standard for PMUs in power systems \cite{martin2014overview} and inverter communication follows the Modbus TCP/IP protocol.%
\begin{table}[thpb]
\begin{center}
\caption{Communicated data in the Controller HIL test setup}
      \label{Tab:CommData}
\begin{tabular}{|  m{2.6cm} | m{1.2cm}| m{1.00cm} | m{1.9cm}|}
 \hline \quad \quad \quad Data & \quad From & \quad To & Comm. Protocol \\ 
\hline \hline
Active and reactive power at 6 points & PMUs 1-6 & Controller & IEEE C37.118 \\ 
\hline Voltage, current, and frequency at 6 points & PMUs 1-6 & Controller & IEEE C37.118 \\
\hline Battery SoC & Inverter & Controller & \quad Modbus \\
\hline PV Generation & Inverter & Controller & \quad Modbus \\
\hline Inverter active and reactive power reference & Controller & Inverter & \quad Modbus \\ \hline
\end{tabular}
\end{center}
\end{table}

 \begin{figure}[thpb]
      \centering
     \includegraphics[width=9cm]{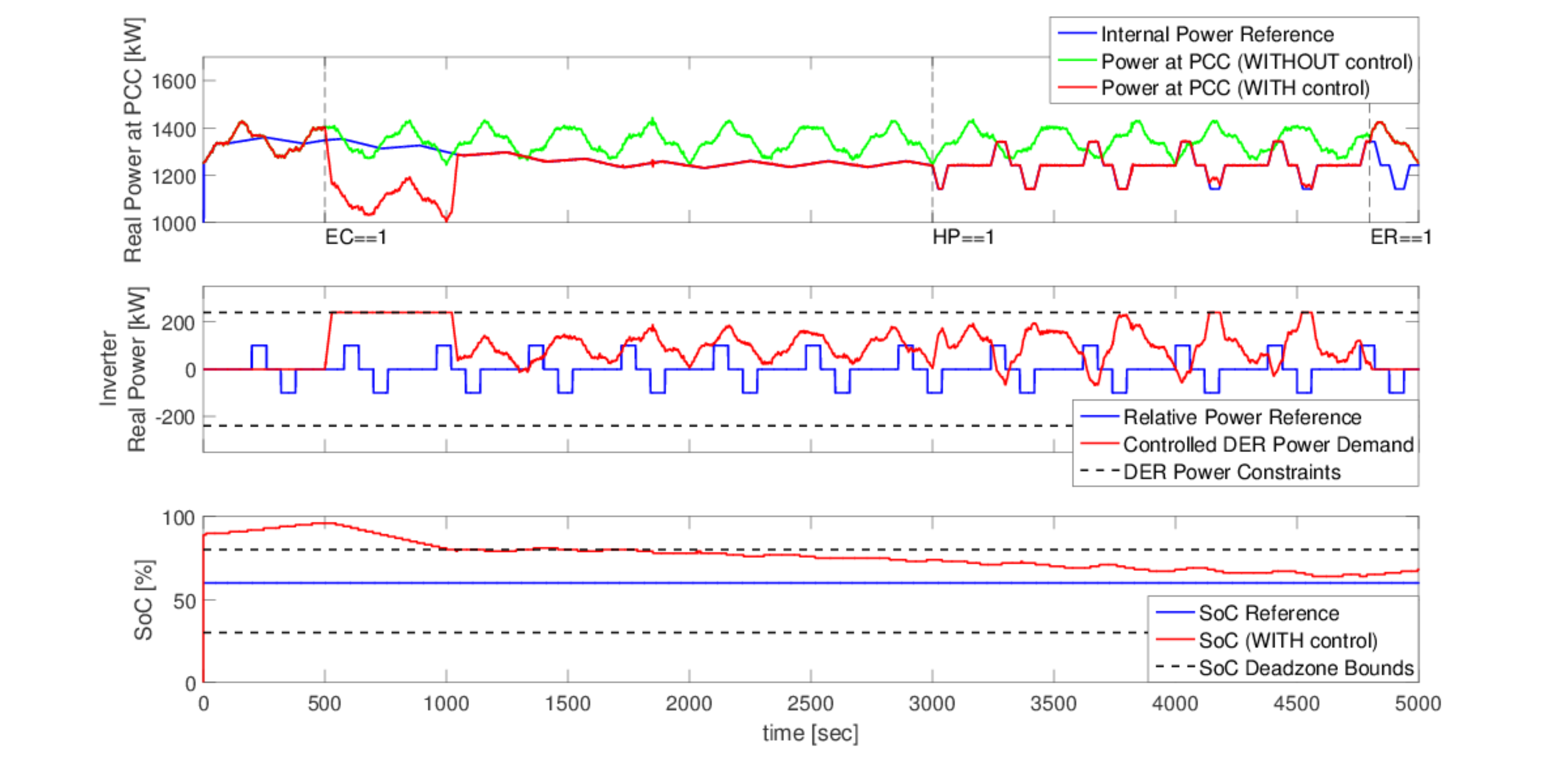}
      \caption{Active Power control at PCC (top), Inverter Input (middle), and SoC variation (bottom)}
      \label{P}
   \end{figure}

 \begin{figure}[thpb]
      \centering
     \includegraphics[width=9cm]{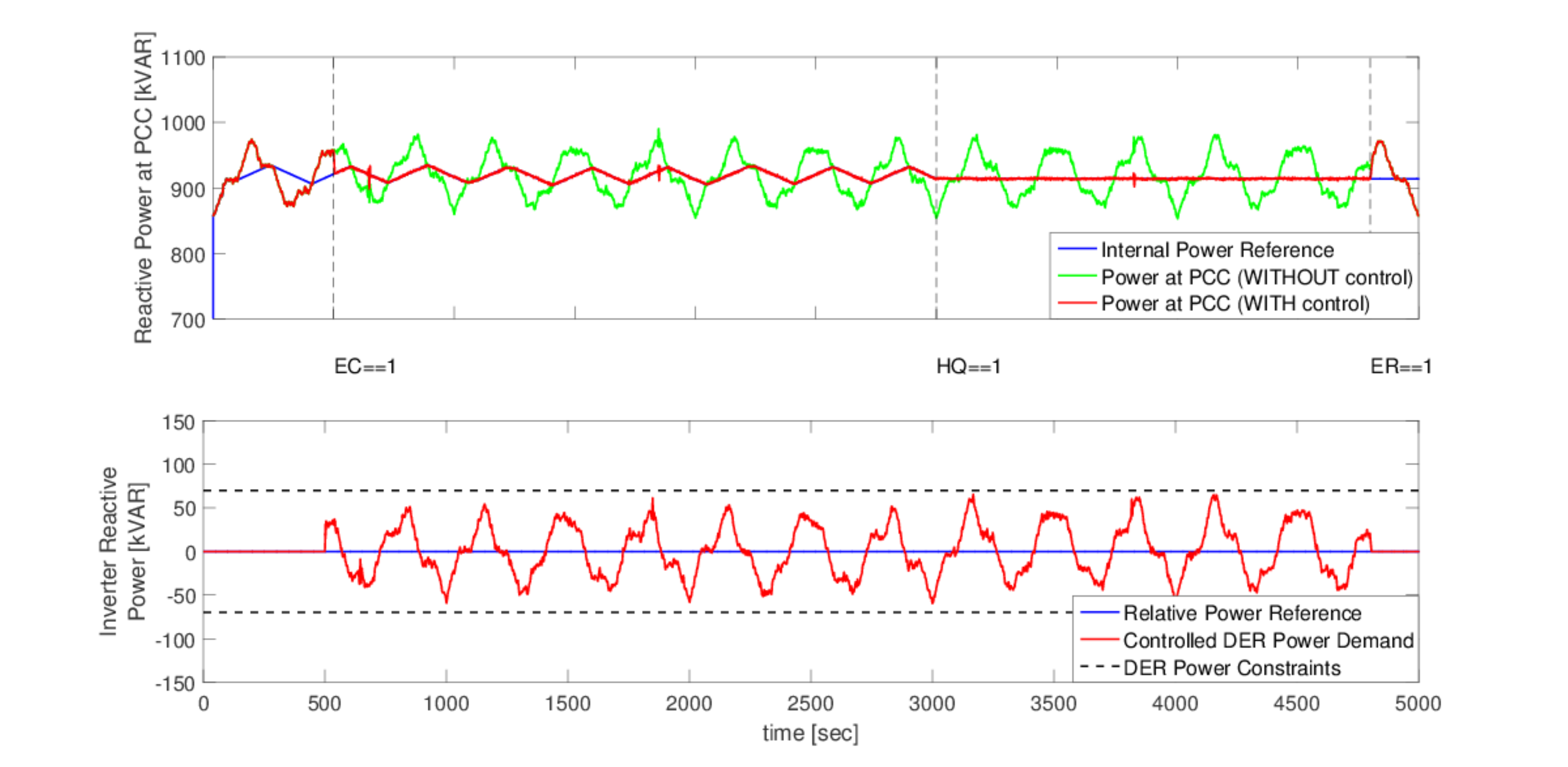}
      \caption{Reactive Power control (top), and Inverter Input (bottom)}
      \label{Q}
   \end{figure}
   
    \begin{figure}[thpb]
      \centering
     \includegraphics[width=8.5cm]{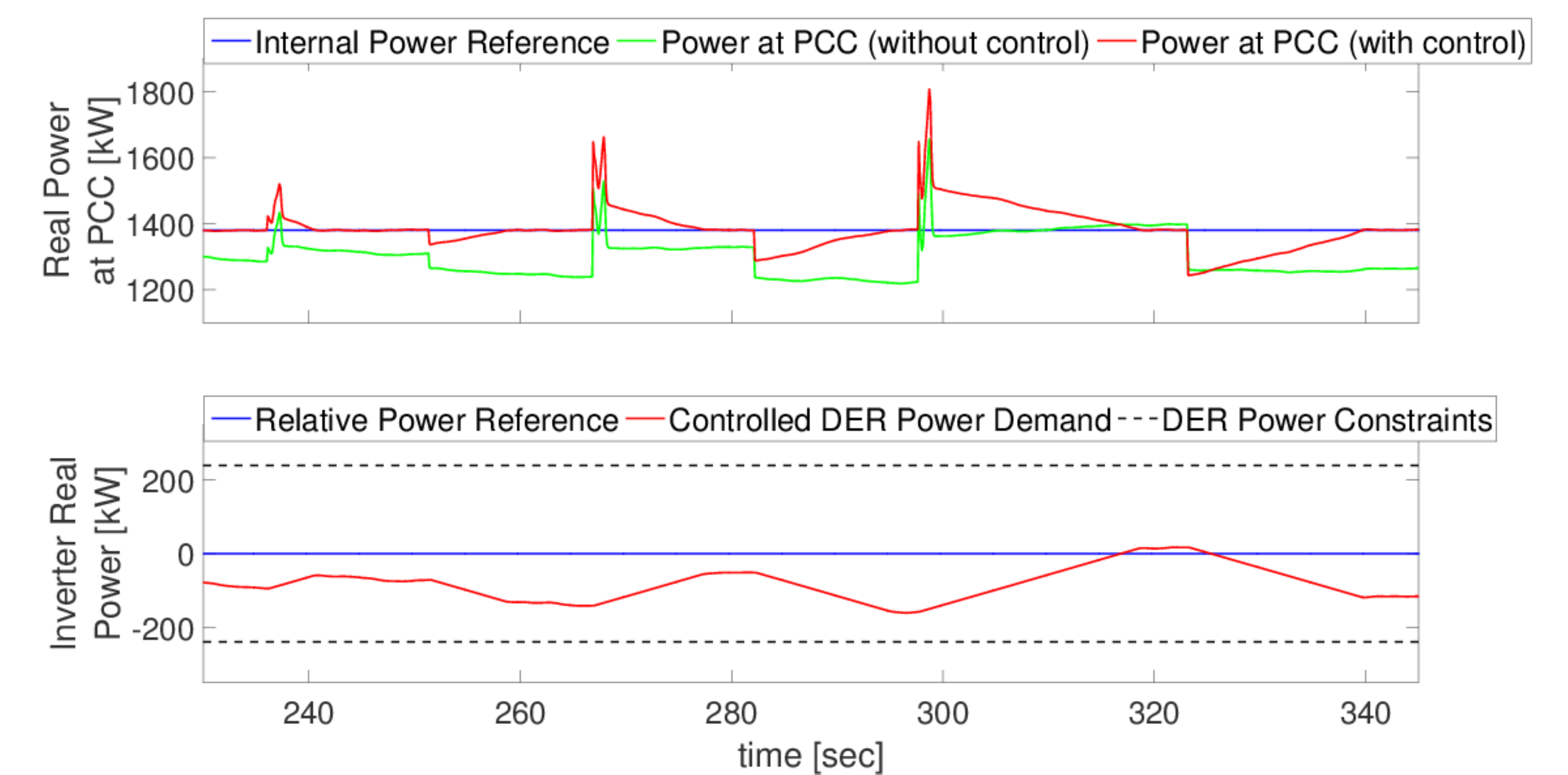}
      \caption{Load Switch Test with Slow Inverter}
      \label{slow}
   \end{figure}
   
    \begin{figure}[thpb]
      \centering
     \includegraphics[width=8.5cm]{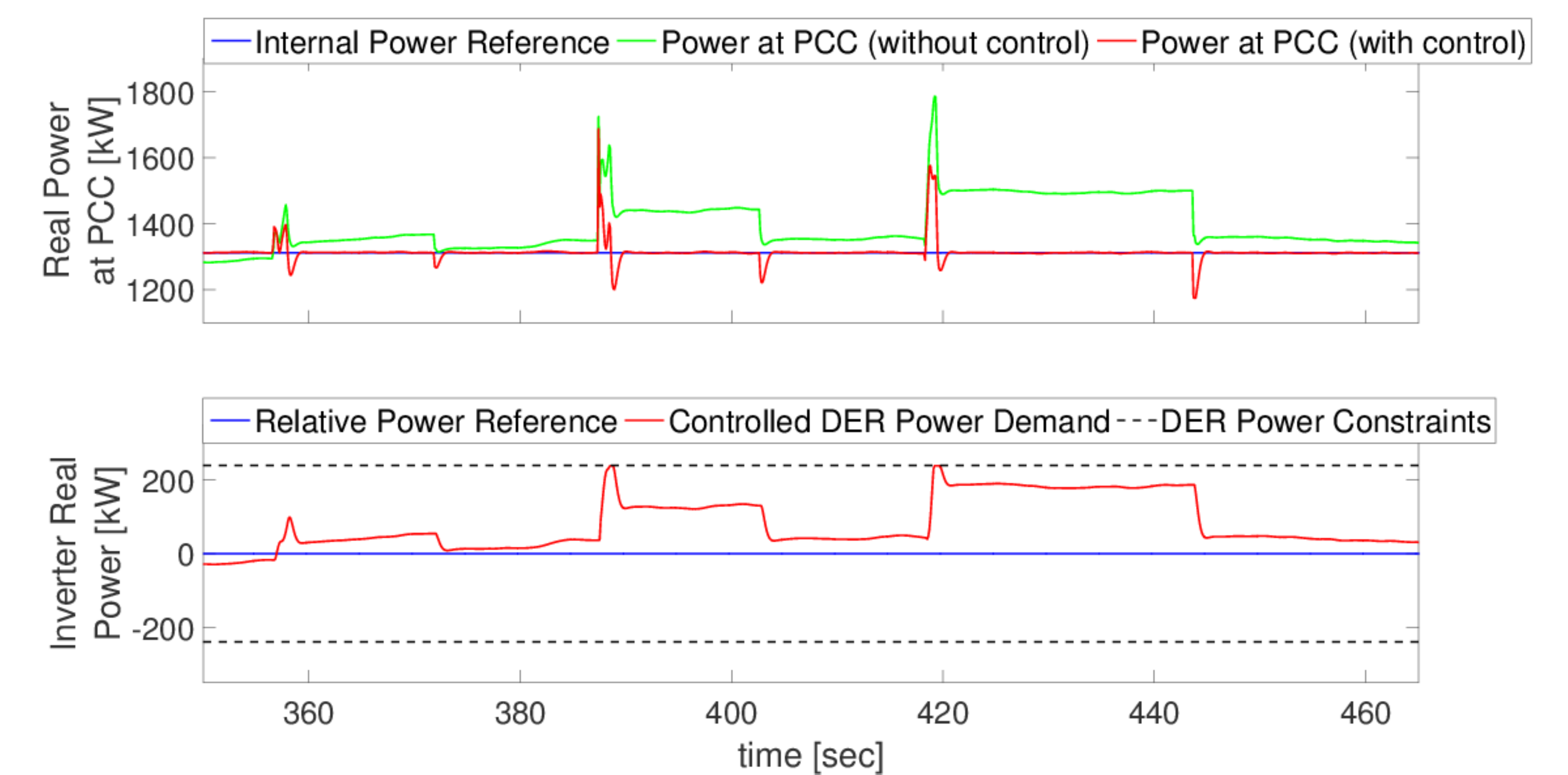}
      \caption{Load Switch Test with Fast Inverter}
      \label{fast}
   \end{figure}
\section{RESULTS}   
Microgrid control and scheduling results are summarized in this section. The two simulated scenarios aim to examine different features of the controller for illustration purposes. 

\subsection{Demand Following Test}
The first test examines power control and tracking at PCC under different PCC power requirements during the total test duration of 5000~seconds. The controller is off during the first 500~seconds, operates in the adaptive reference mode from 500~s to 3000~s, in the manual reference mode from 3000~s to 4800~s, and is commanded to switch off at 4800~s. Additionally, as observed in the SoC plot, the starting SoC of the battery is set outside the dead zone band for test purposes. The controller is activated at 500~s from when it is commanded to operate in the adaptive reference mode. However, by this time, since the SoC has already grown largely out of limits and passed its absolute limits, the only priority of the control system becomes SoC recovery until it reaches the safe zone. This is done by operating the inverter in the full power mode and continues until SoC reaches safe zone at around 1000~s. Afterwards, the controller switches to the normal adaptive reference mode until time 3000~s and the power measured at PCC is able to follow the reference. The reference computed by the adaptive reference computation module is shown by blue in the top figure. Inverter's control input during this period is shown by red in the middle figure and falls within the inverter's power limits. At time 3000~s, the controller is switched from adaptive to non-adaptive reference mode, where the controller is able to follow a trapezoidal reference set by the user. As observed in Figures~\ref{P} (middle) and \ref{Q} (bottom), the inverter's control input is barely reaching its limits after time 1000~s, which means the reference variations are within the inverter's power control capability. The controller is finally switched off at 4800~s.

\subsection{Load Switch Test}
This test is designed to emulate more transient microgrid events. In particular, we have studied abrupt load switching events and the effect of inverter's ramp rate limitation and the communication delay on the controller's ability to suppress those events. The test scenario comprises the microgrid with its usual time-varying load demand while an additional 50~kW motor is suddenly switched in. The switch-in event causes PCC power to experience a sudden jump, however, the controller should be able to recover the previous PCC power level in a timely manner. After successful recovery, the 50~kW motor is switched off and a 100~kW motor is switched at this time. A similar scenario then happens for a 150~kW motor. The tests are performed for two different values of inverter ramp rates. Figure~\ref{slow} shows slow ramp rate power control with a maximum ramp rate of 8~kW/s while Figure~\ref{fast} shows controlled power for a fast inverter with ramp rate of 80~kW/s. In both cases, a controller delay of one time step and a communication delay of one time step exist. The fast inverter obviously outperforms the slow one despite the existing delays in both cases. This is apparent in both PCC power plots (top) and inverter power plots (bottom). In case of the faster ramps (Fig. \ref{fast}), the inverter not only corrects the steady state power level but also partly diminishes the effects of fast power transients that occur during the load switching (apparent in the instantaneous spikes after each event in Fig.~\ref{fast}). This indicates that with the 10~Hz communication frequency and despite delays, the controller is still capable of capturing and controlling  highly transient power fluctuations provided that the inverter is fast enough. 

\section{CONCLUSIONS}
A centralized power scheduling and control system is developed for a microgrid that is simulated in real-time and communicates with the controller over the Internet at a rate of 10~Hz. The controller is able to compute a power reference for the microgrid, adjust power to that reference, and control SoC of the storage unit within the microgrid simultaneously. It is shown that despite the inherent delays in the system, the controller is able to react to both steady state and transient events in a timely manner while maintaining battery State of Charge within its desirable bounds.

\section*{ACKNOWLEDGEMENTS}
This work is associated with the Charge Bliss Renewable Microgrid Project, led by Charge Bliss, Inc., and supported by the California Energy Commission (CEC), under CEC Agreement No. EPC-14-080, a project proposed in response to CEC Solicitation PON 14-301, Demonstrating Secure, Reliable Microgrids and Grid-linked Electric Vehicles to Build Resilient, Low-Carbon Facilities and Communities. The authors would also like to acknowledge Sai Akhil R. Konakalla from UCSD and Charles Wells from OSIsoft for their valuable contributions to this project. 
%%%%%%%%%%%%%%%%%%%%%%%%%%%%%%%%%%%%%%%%%%%%%%%%%%%%%%%%%%%%%%%%%%%%%%%%%%%%%%%%

\bibliographystyle{IEEEtran}
\bibliography{IEEEabrv,mybib}

\end{document}